\documentclass{amsart}

\title{Embeddings of the Standard Model in $E_8$}

\author{Robert Arnott Wilson}

\date{First draft: 20th July 2025. This version: 28th July 2025.}

\address{Queen Mary University of London}
\email{r.a.wilson@qmul.ac.uk}

\newcommand{\RR}{\mathbb R}

\newcommand{\CC}{\mathbb C}
\newcommand{\so}{\mathfrak{so}}

\newcommand{\su}{\mathfrak{su}}
\newcommand{\uu}{\mathfrak{u}}

\newcommand{\rep}{\mathbf}

\begin{document}
\begin{abstract}
I present a modified version of the Manogue--Dray--Wilson `octions' model of elementary particles,
that overcomes some of the objections to that model that have been raised. In particular, I restore the
compactness of the Standard Model 
gauge group, and show how the symmetry-breaking of the weak $SU(2)$ relates to the symmetry-breaking between the three
generations of elementary fermions.

In the process of attempting to implement a Dirac equation for three generations of fermions simultaneously, it turns out that
some parts of the $E_8$ model are not required for the Standard Model, which is entirely contained in the subalgebra
$\so(7,3)$. In particular a re-interpretation of the Dirac spinors allows us to interpret part of the model as quantum gravity,
which is then compared to General Relativity. The general structure of the model shows that the mixing angles depend on masses,
and that the masses emerge from quantum interactions with the dynamic background spacetime (vacuum). Some sample calculations are
given to support these predictions.
\end{abstract}

\maketitle

\section{Introduction}
In \cite{MDW} an embedding of the Standard Model (SM) of Particle Physics into the real Lie algebra
of type $E_{8(-24)}$ was proposed. One of the main objections to this model is that the compact $SU(3)$ colour symmetry group
has been replaced by the split real form $SL(3,\RR)$. It is argued in the paper \cite{MDW} that the complex structure of the spinors allows one
to effectively switch between different real forms, so that this discrepancy is unimportant.

On the other hand, purists would argue \cite{DG} 
that it is of the essence in Yang--Mills theories that the gauge group is compact, and would therefore reject
this model as unphysical. The argument in a nutshell is that gauge bosons must be represented by anti-Hermitian matrices (mathematicians'
convention), and therefore generate a compact Lie group. A non-compact gauge group would have gauge fermions instead, which is normally
regarded as a contradiction in terms. It is not considered physically reasonable for three of the gluons to be bosons and the other five to be fermions.

It is in fact quite straightforward to modify the model to repair this defect. Instead of splitting the group $Spin(7,3)$ into 
$Spin(4)\otimes Spin(3,3)$ and then restricting $Spin(4)$ to $SU(2)$ and $Spin(3,3)$ to $SL(3,\RR)$, we split it as
$Spin(1,3)\otimes Spin(6)$, and restrict from $Spin(1,3)$ to $Spin(3)=SU(2)$ and from $Spin(6)$ to $SU(3)$, thus solving the basic problem.
There are, however, some additional technicalities that need to be addressed, which is the main purpose of this paper.
Before we embark on this, we examine the various options in order to demonstrate that we really have no choice.

\section{The landscape of $E_8$ models}
Amongst the various $E_8$ models that have been proposed recently \cite{MDW,E8,chirality,Chester}, 
there is a consensus that a restriction to $D_8$ is first required,
in order to separate bosons from fermions, and that a further restriction from $D_8$ to $D_5+D_3$ is required in order to separate
the gauge group ($D_5$) from the spin symmetries ($D_3$). There is also a consensus that the required real form of $E_8$ is $E_{8(-24)}$,
which contains $Spin(12,4)$ as the appropriate real form of $D_8$.

On the other hand,
there is no consensus about the appropriate real form of the splitting into $D_5+D_3$ among the five available options:
\begin{align}
Spin(10) & \otimes Spin(2,4)\cr
Spin(9,1) & \otimes Spin(3,3)\cr
Spin(8,2) & \otimes Spin(4,2)\cr
Spin(7,3) & \otimes Spin(5,1)\cr
Spin(6,4) & \otimes Spin(6)
\end{align}
We can surely eliminate the third case, because $Spin(8,2)$ does not contain the gauge group $SU(3) \times SU(2)$
of the nuclear forces, and we can eliminate the last one, because
$Spin(6)$ does not contain the Lorentz group $Spin(3,1)$. 

The other three cases all contain $SU(3)\times SU(2)$ via a splitting of $D_5$ into
$D_3+D_2$:
\begin{align}
Spin(6)\otimes Spin(4) & \subset Spin(10)\cr
Spin(6)\otimes Spin(3,1) & \subset Spin(9,1)\cr
Spin(6)\otimes Spin(1,3) & \subset Spin(7,3).
\end{align}
The first case here is the original Georgi $SO(10)$ GUT from 1974, combined with Penrose twistors \cite{twistors,twistorlectures}
acted on by $Spin(2,4)$, as an extension of
the Dirac spinors. In some sense this is the `obvious' case to look at, but after 50 years this model has not lived up to its initial promise.
The second case is studied in \cite{Chester}, where the group $Spin(3,3)$ is used to model three generations, by extending the energy coordinate from
one to three degrees of freedom. The $Spin(7,3)$ case is studied in \cite{MDW}, but with a different real form of the splitting $D_3+D_2$, that is
not consistent with the SM:
\begin{align}
Spin(3,3) \otimes Spin(4) \subset Spin(7,3).
\end{align}

The key to choosing the correct option is to choose an appropriate copy of $U(1)$ 
to define a complex structure.
In the $SO(10)$ GUT, a choice of $U(1)$ breaks the symmetry to $U(5)$ and reduces to the Georgi--Glashow $SU(5)$ GUT \cite{GG}. 
But this model predicts
proton decay, which has never been observed, so we can rule out this case. In the other two cases, $U(1)$ can only act on $8$ of the $10$ real
coordinates, leading to the groups
\begin{align}
U(4) \otimes Spin(1,1) & \subset Spin(9,1)\cr
U(3,1) \otimes Spin(1,1) & \subset Spin(7,3)
\end{align} 
Both of these can be considered as variants of the Pati--Salam model \cite{PatiSalam}, in which the three-colour $SU(3)$ symmetry is extended to
include a fourth colour for leptons. The first is closer to the original, in which there are gauge bosons that mix the leptons and quarks.
Such mixing is not observed, however, so this model can be ruled out also. 

In the second case, on the other hand, there are no such gauge bosons,
and the qualitative distinction between leptons and quarks is maintained. Indeed, $SU(3,1)$ makes a clear distinction between three
`spacelike' colours for quarks, and one `timelike' colour for leptons.
However, the embedding of $SU(3,1)$ in $Spin(7,3)$ does not really support this four-colour interpretation. Instead it supports a splitting into
$7$ spacelike `colours', and three timelike `generations'. The symmetry-breaking down to $SU(3,1)$ then breaks the colour symmetry to $1+2\times 3$,
and breaks the generation symmetry to $1+2$.

This gives us an appropriate 
choice of electromagnetic $U(1)$, which splits in the Standard Model into a hypercharge $U(1)_Y$ and the third component of
weak isospin. The latter breaks the symmetry of $Spin(3)$ down to $Spin(2)$, while the former breaks the symmetry of $Spin(7)$ down to $U(3)$.
The centralizer of $U(1)_Y$ is
\begin{align}
U(3) \otimes Spin(1,3)
\end{align}
whose compact part is exactly the gauge group of the Standard Model.
We also have $Spin(1,3)=SL(2,\CC)$ as the complexification of $Spin(3)=SU(2)$, which is what is used in the SM to give mass to the
mediators of the weak interaction. 

At this point we have completed our survey of the landscape, and come to the conclusion that there is
a unique physically plausible embedding of the Standard Model in $E_{8(-24)}$. We even have a choice between regarding hypercharge as fundamental,
in which case we obtain the unbroken gauge group of the SM, including a complex gauge group for the weak force,
or regarding charge as fundamental, in which case the symmetry group
is instead a complex scalar plus $SU(3,1)$.

\section{Variants of the Dirac algebra}
It is usual in particle physics to regard hypercharge as more fundamental than charge, but the opposite is true in classical physics.
In the interests of explaining the emergence of classical physics from quantum physics, or of quantisation of gravity, or unification,
it may therefore be more
useful to switch back to thinking of charge as more fundamental, and regarding the charged electrons 
as being fundamentally different from the uncharged neutrinos. In this case, we need to investigate the structure of $Spin(7,3)$
from the point of view of the subgroup $SU(3,1)$, rather than the SM gauge group.

It is straightforward to calculate the splitting of the adjoint representation of $Spin(7,3)$ under $SU(3,1)$. In addition to the $15$-dimensional
adjoint of $SU(3,1)$ itself, there are $15$ complex dimensions, which split into complex irreducibles as $4+6+4+1$. On further restriction to
$SO(3,1)$, the $4$s become Lorentz vectors, and the $6$ becomes a Lorentz adjoint representation. In other words, we have a representation
of the Lorentz group looking very much like the complex Dirac algebra without the scalars. However, the algebraic structure of the
Lie algebra $\so(7,3)$ is very different from the algebraic structure of the complex Clifford algebra $\CC \ell(3,1)$.
In particular, the complex scalars of $\CC\ell(3,1)$ are replaced in $\so(7,3)$ by the adjoint representation of $SU(3,1)$,
which
contains $9$ gauge bosons, that are identified in the SM as the photon plus $8$ gluons.
It also contains $6$ gauge fermions, that have no electric charge, and must therefore be three neutrinos and three antineutrinos. Because the
Dirac equation has to be written in terms of the quotient space $\so(7,3)/\su(3,1)$, these neutrinos have zero Dirac mass, although they could conceivably have
some other type of mass.

With this interpretation, we have identified the algebra $\su(3,1)$ with a complete set of particles that have zero Dirac mass, and we can interpret
these particles as forming a quantum vacuum. The Dirac algebra on top of this extends this basic vacuum to include pair creation and annihilation as well.
However, these particles have no momentum or energy at this stage, so they must acquire these properties from somewhere else, perhaps
from the remaining group $Spin(5,1)$. The adjoint representation of the latter 
splits as $4+6+4+1$ real representations under the Lorentz group $Spin(3,1)$. Thus we have another
type of Dirac algebra here, this time real rather than complex.

There is therefore a big question as to what, if anything, is the relationship between these two `Dirac algebras', and what, if anything, is the relationship
between the Lorentz group $SO(3,1)$ inside $Spin(7,3)$, and the Lorentz group $Spin(3,1)$ inside $Spin(5,1)$? 
Is there an abstract definition of momentum and energy in $Spin(3,1)$ independent of matter inside $Spin(7,3)$, or can we only define momentum
and energy in relation to the ambient matter, using the subgroup $SO(3,1)$ of $Spin(7,3)$? Common sense says we can only define momentum with
respect to matter, rather than intrinsically, which implies that $Spin(5,1)$ has nothing to do with momentum, and therefore has nothing to
do with the Dirac equation.

To understand these questions
better, and perhaps answer them, we need to do some detailed calculations. We use the same notation as in \cite{MDW}, in which the $16$
coordinates of the natural representation of $SO(12,4)$ are labelled by $8$ octonions $1,i,j,k,l,il,jl,kl$ and $8$ split octonions $U,I,J,K,L,IL,JL,KL$.
Generators for the individual copies of $SO(2)$ and $SO(1,1)$ acting on two 
coordinates are called $X_{aB}$ if $a$ is an octonion and $B$ is a split octonion,
and $D_{a,b}$ or $D_{A,B}$ if both coordinates are of the same type. Occurrences of $1$ and/or $U$ are omitted in the notation if no confusion 
can result.

We keep the same complex structure from \cite{MDW}, defined by $X_1$ acting on the $2$-space spanned by $1$ and $U$, but move $Spin(3,1)$ to $I,J,K,L$,
so that colour $SU(3)$ can act on $i,j,k,il,jl,kl$ as three complex coordinates, and weak $SL(2,\CC)$ acts on $l,IL,JL,KL$.
Effectively, then, $l$ represents the hypercharge, and $IL$, $JL$ and $KL$ can be identified as the three components of weak isospin.
Alternatively, we may wish to choose our notation so that $IL+JL+KL$ is the conventional third component of weak isospin, so that 
permutations of $IL$, $JL$ and $KL$ can induce generation symmetries on the elementary fermions. Either way, the first two components of weak
isospin are used for the generation symmetry.

Generators for $Spin(5,1)$ acting on the six coordinates $1,U,I,J,K,L$ can be chosen analogous to the Dirac matrices as follows:
\begin{align}
\gamma_1 &\rightarrow D_I,\cr
\gamma_2 &\rightarrow D_J,\cr
\gamma_3 &\rightarrow D_K,\cr
\gamma_0 &\rightarrow D_L,\cr
i\gamma_5 &\rightarrow X_1.
\end{align}
In particular, there is no scalar $i$, and no analogue of $\gamma_5$, only the pseudoscalar $i\gamma_5$ (which is nevertheless a 
Lorentz scalar, so could still be used for the mass term in a variant of the Dirac equation). The Lorentz transformations themselves are generated by
$D_{I,L}$, $D_{J,L}$ and $D_{K,L}$, and linear combinations of these. The grading of the Clifford algebra can be expressed as follows:
\begin{align}
1: & \quad D_I, D_J, D_K, D_L\cr
2: &\quad  D_{I,J}, D_{J,K}, D_{K,I}, D_{I,L}, D_{J,L}, D_{K,L}\cr
3: & \quad X_I, X_J, X_K, X_L\cr
4: & \quad X_1
\end{align}
It is also possible to swap the degree $1$ and degree $3$ parts of the algebra, if desired.

Similar generators for $Spin(7,3)$ can be obtained by choosing a suitable copy of $SO(3,1)$ inside $SU(3,1)$, for example choosing
Lorentz scalars $l$ and $KL$, and Lorentz vectors
\begin{align}
&i,j,k,IL\cr
& il,jl,kl,JL
\end{align}
or something similar. 
In order to lay bare the structure, we add first the pseudoscalars (degree $4$), then the Lorentz adjoints (degree $2$), and finally the
Lorentz vectors, to obtain a series of Lie algebras as follows:
\begin{align}
\su(3,1) \subset \uu(3,1)+\so(1,1) \subset \so(6,2)+\so(1,1) \subset \so(7,3)
\end{align}
The pseudoscalars (Lorentz scalars, representing the Higgs boson) can be taken as
\begin{align}
&X_{lKL}\cr
& D_{i,il}+D_{j,jl}+D_{k,kl} +D_{IL,JL}
\end{align}
The choice of pseudoscalar determines which copy of $SU(3,1)$ we are using, but is otherwise unimportant.

The adjoint representation of our chosen copy of $SO(3,1)$ inside $SU(3,1)$ is then spanned by
\begin{align}
\label{newso31}
&D_{i,j}+D_{il,jl}, D_{j,k}+D_{jl,kl}, D_{k,i}+D_{kl,il}, \cr
&X_{iIL}+X_{ilJL}, X_{jIL}+X_{jlJL},X_{kIL}+X_{klJL}.
\end{align}
One of the two other copies is obtained from this by changing the signs to 
\begin{align}
\label{oldso31}
&D_{i,j}-D_{il,jl}, D_{j,k}-D_{jl,kl}, D_{k,i}-D_{kl,il}, \cr
&X_{iIL}-X_{ilJL}, X_{jIL}-X_{jlJL},X_{kIL}-X_{klJL}.
\end{align}
The third copy can be taken as
\begin{align}
&D_{i,jl}-D_{j,il}, D_{j,kl}-D_{k,jl}, D_{k,il}-D_{i,kl},\cr
& X_{ilIL}+X_{iJL}, X_{jlIL}+X_{jJL}, X_{klIL}+X_{kJL}. 
\end{align}

Finally, the Lorentz vectors are
\begin{align}\label{fourvectors}
&D_{i,l}, D_{j,l}, D_{k,l}, X_{lIL};\cr
& D_{il,l}, D_{jl,l}, D_{kl,l}, X_{lJL};\cr
& X_{iKL}, X_{jKL}, X_{kKL}, D_{IL,KL};\cr
& X_{ilKL}, X_{jlKL}, X_{klKL}, D_{JL,KL}.
\end{align}
The first two have natural signature $(1,3)$, and together form a complex $4$-vector representation of $U(3,1)$.
Either of them, or any linear combination, could be chosen to represent $\gamma_1$, $\gamma_2$, $\gamma_3$ and
$\gamma_0$. The other two have signature $(3,1)$, and are obtained by applying $\gamma_5=X_{lKL}$.
They form another complex $4$-vector for $U(3,1)$.

Of course, the Lie algebra $\su(3,1)$ is itself a $15$-dimensional algebra, which can also be generated by analogues of the 
Dirac matrices, since it is just another real form of $D_3$. But this is the only real form that cannot be obtained just by multiplying some of the
Dirac matrices by $i$, and therefore a more subtle construction is required.
In place of $\gamma_1$, $\gamma_2$ and $\gamma_3$ we take (anti-Hermitian) Gell-Mann matrices on the last three coordinates,
so that they generate $\su(3)$, and the Lie brackets of pairs of them generate
a copy of $\so(3)$ inside it:
\begin{align}
\kappa_1:=\begin{pmatrix}0&0&0&0\cr 0&0&i&0\cr 0&i&0&0\cr 0&0&0&0\end{pmatrix}, \quad
\kappa_2:=\begin{pmatrix}0&0&0&0\cr 0&0&0&0\cr 0&0&0&i\cr 0&0&i&0\end{pmatrix}, \quad
\kappa_3:=\begin{pmatrix}0&0&0&0\cr 0&0&0&i\cr 0&0&0&0\cr 0&i&0&0\end{pmatrix}.
\end{align}
In place of $\gamma_0$ we take a Hermitian matrix mixing the time and space coordinates:
\begin{align}
\kappa_0:=\begin{pmatrix}0&i&i&i\cr -i&0&0&0\cr -i&0&0&0\cr -i&0&0&0\end{pmatrix}, 
\end{align}
so that the products $[\kappa_0,\kappa_1]$, $[\kappa_0,\kappa_2]$ and $[\kappa_0,\kappa_3]$ generate Lorentz transformations to
extend the algebra to $\so(3,1)$, represented by the real matrices.

In terms of the $E_8$ notation of \cite{MDW}, these matrices can be written as
\begin{align}
\kappa_1=D_{i,jl}+D_{j,il},\quad \kappa_2=D_{j,kl}+D_{k,jl},\quad \kappa_3= D_{k,il}+D_{i,kl}, 
\end{align}
as generators for $\su(3)$, extended by
\begin{align}
\kappa_0= X_{(i+j+k)IL} - X_{(i+j+k)lJL}
\end{align}
to generate $\su(3,1)$.
These four matrices 
act on the two $4$-vectors  listed in (\ref{fourvectors}), which are swapped by $\kappa_5:=X_{lKL}$. 
Now the copy of $\so(3,1)$ generated by the Lie brackets of these $\kappa$ matrices is exhibited in (\ref{newso31}),
compared to the Standard Model copy which is the one exhibited in (\ref{oldso31}).
The difference lies only in the signs, so is rather subtle, but both mathematically and physically very important.

The change of sign means that spinors are not required for describing spin. Instead, classical rotation using position and momentum vectors
is sufficient. The physical reason for this is that an electron in isolation requires a spinor for its description, but an electron is never in isolation,
it is always connected to the rest of the universe by electromagnetic and gravitational forces. 
Classical rotating objects that are connected, say by a piece of string, to the rest of the universe, also have the spin $1/2$ property, that after
a $360^\circ$ rotation the string becomes twisted, but after a $720^\circ$ rotation, the string is again untwisted.
Hence there is no physical or mathematical reason why spinors are required for describing spin.

The Dirac
equation is an equation that relates mass to spin, so our replacement must relate mass to the position and momentum $4$-vectors,
using the $\kappa$ matrices that generate $\su(3,1)$, and represent the spacetime background, or vacuum.
The exact formula required is far from obvious, and I make no attempt to guess what it is. Instead, let us first look at experimental
evidence to guide us towards the correct equation. The basic principle is that mass should be related to rotation with respect to the
background, and therefore, by duality or by Mach's Principle, to rotation of the background with respect to the particle concerned.
Moreover, since the vectors are complex, we need to consider both mass and charge, and momentum and current.

We therefore aim for a modified Dirac equation using the $\kappa$ matrices, that generate $SU(3,1)$,
 instead of the $\gamma$ matrices that generate  $Spin(4,1)$.  The Dirac spinor (four complex coordinates, divided into two `left-handed' and two
`right-handed') is replaced by this pair of complex vectors (eight complex coordinates, divided into four `left-handed' and
four `right-handed'). To reduce to the standard Dirac equation, we must reduce from $4$-vectors to $2$-vectors by first splitting each
$4$-vector into two $2$-vectors, so the group reduces to 
$SU(1,1)\times SU(2)$. Then we stack the two halves of each $4$-vector on top of each other, so they are acted on partly 
by $SU(1,1)$ and partly by $SU(2)$. Then we regard the two $4$-vectors as being complex scalar multiples of each other, so that we can combine
the two groups $SU(1,1)$ and $SU(2)$ into $SL(2,\CC)$. Finally, we (mis)interpret the latter as a double cover of the Lorentz group.

This process leaves us with some Dirac spinors that no longer have a reasonable physical interpretation,
which leads to the enduring problem of interpretation of quantum mechanics.
On the other hand, the complex $4$-vectors split into a $2$-vector on complex $(z,t)$ space, defining momentum and current, and a complex
$2$-vector on $(x,y)$ space, defining spin of mass and charge. In particular, spin is described by $SU(2)$ acting on a complex $2$-space,
and is not related to rotation $SO(3)$ acting on real $3$-space.

The new formalism significantly simplifies the Dirac equation, since it no longer requires the undefined concept of spinors,
and at the same time extends from one generation to three.
At first glance the structure looks much more complicated, since in place of the
complex scalar we acquire an entire `gauge group' $SU(3,1)$.
But this means that this generalised Dirac algebra not only contains an analogue of $\gamma_5$ for implementing the weak force, but also contains
a complete $SU(3)$ for implementing the strong force. Since it also contains a generation symmetry in the first two components of weak isospin,
it should in principle support all of the calculations required in the Standard Model.

\section{Numerical predictions}
If this is the case, then it is not actually necessary to use the remaining six coordinates at all, but simply use the algebra $\so(7,3)$
for everything. In other words, our attempt to build an $E_8$ model containing the Standard Model has led to a much smaller
$D_5$ model that does the same thing. 
It is not our purpose here to carry out a complete translation of the Standard Model into the language of $\so(7,3)$, but rather to
provide a `proof of concept' that $\so(7,3)$ contains all the necessary ingredients.
In particular, we shall show in this Section that this algebraic structure allows us to calculate some of the mixing angles of the SM.

First we project the three lepton generations onto the first two components of weak isospin, where we can interpret the 
three generations of electron as the vertices of an equilateral triangle, embedded in the circle $SO(2)=U(1)$. 
There is then a well-defined mass direction, and we can calculate the
angle $\theta$ between this direction and one of the sides of the triangle from the equation
\begin{align}
\frac{\cos(60^\circ-\theta)}{\cos\theta} = \frac {m(\tau)-m(e)}{m(\tau)-m(\mu)},
\end{align}
which yields an angle
\begin{align}
\theta \approx 33.024^\circ.
\end{align}
Clearly this is an important angle in the lepton-mixing matrix (that is, the PMNS \cite{Pontecorvo,MNS} matrix), so it is no surprise that it is in
agreement with the experimental value of the electron/muon neutrino mixing angle.

To take another example, we embed the four fundamental fermion types (that is, neutrino, electron, up and down quarks)
as the vertices of a square in the charge/hypercharge plane, suitably scaled, in the cyclic order $\nu$, $e$, $d$, $u$. In this case the embedding
is into the group $SO(1,1)$, which is geometrically a hyperbola, defined by the equation
$x^2-y^2=1$. The two branches of the hyperbola distinguish the leptons, which are asymptotic to one branch,
from the quarks, which are asymptotic to the other. We then calculate the
angle between the charge direction and the hypercharge direction from the equation
\begin{align}
\tan \phi = 3/2,
\end{align}
which yields
\begin{align}
\sin^2(\phi/2) & = 1/2 - 1/\sqrt{13}\cr
& \approx .22265.
\end{align}
It is again no surprise that this angle $\phi/2$ is the same as the experimental value of the weak mixing angle.

For our third example, we combine these two planes into the $4$-space spanned by $l, IL, JL, KL$, and for clarity of the generation symmetry
we choose $l$ and $IL+JL+KL$ for the charge/hypercharge plane, so that $IL-JL$, $JL-KL$ and $KL-IL$ denote three generations of
uncharged particles (neutrinos or antineutrinos), and $IL+JL$, $JL+KL$, $KL+IL$ denote three generations of charged particles,
which, following \cite{Jansson}, I take to be electrons. Using charge and $IL, JL, KL$ as coordinates, for transparency, we then have
\begin{align}
e& = (-1,1,1,0)\cr
\mu & = (-1,0,1,1)\cr
\tau & = (-1,1,0,1)
\end{align}
To span the whole $4$-space we need another particle, say the proton, which interacts equally with all flavours of neutrinos, so must lie in the IL+JL+KL direction, which suggests
\begin{align}
p = (1,1,1,1).
\end{align}
We can then calculate the neutral combination
\begin{align}
e+\mu+\tau+3p = (0,5,5,5)
\end{align}
which can only reasonably be interpreted as five neutrons.

Remarkably, it is true that the total mass of these 6 particles is indeed equal to the mass of five neutrons, to within the bounds of experimental
uncertainty \cite{remarks}. Therefore we can use this equation as a prediction for a more accurate value of the tau mass than is known from
experiment to date:
\begin{align}
m(\tau)_p = 1776.841464(4) \mbox{ MeV}/c^2.
\end{align}
Moreover, this mass equation provides some clues and constraints as to how to build a Dirac equation that relates quantum numbers of
elementary particles to continuous approximations to the quantum background.

Comparing this third example with the first suggests that it should be possible to find the rest of the mixing angles as mass ratios of suitable
linear combinations of these five fundamental masses. Perhaps the most obvious mass ratios to consider are those between electron, proton and
neutron, which we arrange with equal charge in numerator and denominator, as follows:
\begin{align}
\cos\psi & = \frac{m(e)}{m(n)-m(p)}\cr & \approx .3952\cr
\cos\chi & = \frac{m(p)}{m(n)-m(e)}\cr &  \approx .999167.
\end{align}
These equations yield angles
\begin{align}
\psi & \approx 66.72^\circ\cr
\chi &\approx 2.339^\circ
\end{align}
both of which can be found in the CKM \cite{Cabibbo,KM} matrix, as the two angles that fix the first generation of quarks.
The first agrees with the overall CP-violating phase, while the second is the mixing angle between the second and third generations.

The Cabibbo angle, mixing the first two generations of quarks, is almost exactly $20^\circ$ less than the corresponding angle in the PMNS matrix,
calculated above as $33.024^\circ$. This looks like a ridiculous coincidence, until we consider the possibility that the generation symmetry for quarks cubes to the colour symmetry rather than cubing to the identity. This hypothesis introduces rotations of order $9$ into the formalism, and it is easy to
imagine that a $20^\circ$ rotation arises from the difference between a $4/9$ rotation and a $1/2$ rotation. It may also be relevant that the
PMNS matrix contains a mixing angle that is experimentally indistinguishable from $50^\circ$, which is the difference between a $1/4$ rotation
and a $1/9$ rotation. If so, then we should also expect to see the difference between $7/9$ and $3/4$, that is $10^\circ$, appearing somewhere.
This is close to the electron/tau neutrino mixing angle of around $8.5^\circ$, although we must look elsewhere for the difference of around $1.5^\circ$
between these two values.

These examples between them potentially deal with $5$, or even $6$, of the $9$ mixing angles in the SM, and demonstrate that the proposed $\so(7,3)$ model
may be able to explain a number of currently unexplained features of the SM.
Since the algebra also contains a copy of the Lorentz group, in such a way that the time/energy coordinate mixes with $IL, JL, KL$, the mixing
angles calculated above will also be affected by the energy. Thus the angles are rotation-invariant, but not Lorentz-invariant.
This `running' of mixing angles with energy is indeed a feature of the Standard Model.

There is certainly enough complexity in the $4$-space spanned by $l,IL,JL,KL$ to contain the remaining mixing angles,
but to find them explicitly we may need a more rigorous description of the quantum numbers, and the full Dirac equation.
The quaternionic quantum numbers described in \cite{Jansson} may be useful in this regard, but we may also need to relate the
spin $SU(2)$ symmetry, acting on $j,k,jl,kl$, to the generation (or weak isospin) symmetry group $SO(3)$, acting on $IL,JL,KL$.

\section{Relativity and quantum gravity}
In the Standard Model there is some ambiguity as to what is meant by `the' Lorentz group, since the degree
$2$ part of the complex Clifford algebra splits into the sum of two copies of the Lie algebra $\so(3,1)$. The corresponding groups in
$Spin(6,2)$ are two commuting copies of $Spin(3,1)$, which combine into a copy of $SO(3,1)$ that embeds in $SU(3,1)$. 
But because the spinors have no physical meaning in our model, these copies of $Spin(3,1)$ can no longer be identified
with the Lorentz group. 
This removes the ambiguity, and leaves us with a single copy of
$SO(3,1)$ that does not have any spin representations at all, but is used in the theories of relativity \cite{thooft,GR1,GR2} to describe the
classical forces of electromagnetism and gravity.

It therefore makes sense to investigate the embedding of $SO(3,1)$ in $SU(3,1)$ from the relativistic point of view.
The $9$ dimensions of $SU(3,1)$ outside $SO(3,1)$ form the spin $(1,1)$ representation of $SO(3,1)$, which is the same representation
that is used in GR for the Einstein tensor, and the non-scalar parts of the stress-energy tensor and the Ricci tensor.
What is different here is that instead of the conventional embedding of $SO(3,1)$ in $GL(4,\RR)$ that expresses the principle of general
covariance, we have an embedding in $U(3,1)$. The difference between these two cases is the difference between the signature
$(10,6)$ of $GL(4,\RR)$ and the signature $(6,10)$ of $U(3,1)$.

In a sense, it is possible to convert from one to the other simply by multiplying by $i$. This 
seemingly innocuous change has essentially no effect on the linear parts of the
theory, but rather drastic effects on the non-linear parts. It completely re-writes the symmetry groups,
causes havoc in the definition of general covariance, and changes the sign of the self-interaction of the
gravitational field. But it has essentially no effect on the first-order predictions of the theory,
so it is quite difficult to test the differences between the two models.
In particular, a version of GR that uses $U(3,1)$ instead of $GL(4,\RR)$ cannot be immediately ruled out on experimental grounds.

Conversely, the main argument in favour of using $U(3,1)$ is that $SU(3,1)$ can be quantised in terms of massless
elementary particles, as we have shown, whereas it is known that
quantising $SL(4,\RR)$ leads to a theory that is not renormalizable \cite{GL4R1,GL4R2}. 
Moreover, since the signature is reversed, the notions of fermions and bosons are
swapped, leading to predictions of `supersymmetric partners' of the known particles. The lack of experimental detection of such partners
is another clue that $SU(3,1)$ may be physically viable while $SL(4,\RR)$ is not.

The structure of $Spin(7,3)$ also shows that the proposed group $SU(3,1)$ of spacetime transformations is `emergent' from the
particle interactions described by the Dirac algebra, and does not need to be added in separately. Indeed, we can interpret
$SU(3,1)$ as a description of the local vacuum, on top of which the Dirac algebra describes 
the effect of the local forces. But because $\so(7,3)$ is a simple Lie algebra,
it is impossible to separate the Dirac algebra from the vacuum, and each affects the other through elementary particle interactions.

This process works in both directions, so that not only is the local spacetime symmetry group emergent from the elementary
particle interactions, but also the speed at which elementary particle interactions take place depends on the local spacetime symmetry group.
This is indeed a prediction of GR, that has been tested and found to be correct.
A consequence is that electromagnetic measurement of time using atomic clocks must be corrected for this effect, as indeed it is.

The most obvious non-electromagnetic experimental effect 
that can potentially be explained in this way is the neutron lifetime anomaly. Measurements of the neutron lifetime
in beam experiments and bottle experiments lead to discrepancies of the order of $1\%$ between the two methods, far more than can be
realistically attributed to experimental uncertainties or systematics. Beam neutrinos experience a flat spacetime, as each one only lives in the experiment
for a fraction of a second. Bottle neutrinos, on the other hand, are confined in the bottle for something of the order of 
their entire lifetime, during which time the bottle has rotated
in the gravitational field by about $1/100$ of a complete rotation, on average. The resulting curvature in their experience of spacetime may therefore be
sufficient to explain the $1\%$ discrepancy in the two different measurements.

Other experiments that indicate a connection to the quantisation of spacetime come from anomalous measurements of the lifetimes of
other radioisotopes,
some of which appear to exhibit fluctuations related to solar neutrino activity. Since these neutrinos lie in the group $SU(3,1)$, they alter the
local shape of spacetime, and therefore affect lifetimes in the same way as in the neutron experiments.

\section{Background dependence}
The role of the putative extended
Dirac equation in relativity is to use the new $\kappa$ matrices to unite $\so(3,1)$ with the Einstein tensor in $\su(3,1)/\so(3,1)$, and 
hence to define a unified `mass' that works in both electromagnetism and gravity. But it also shows that this unified mass is dependent on the
structure of the local (quantised) spacetime background, and that the equivalence of electromagnetic and gravitational mass is only local,
and dependent on the background. This is indeed what Einstein's equivalence principle actually says, although it is widely misunderstood
as saying that there is a \emph{universal} concept of mass defined by the Dirac equation. 

The existence of such a universal mass contradicts
the principle of relativity, and is therefore inconsistent. The only way that the Standard Model can insist on universal mass parameters is by
defining a `standard' background vacuum state, which in practice appears to be the average vacuum state experienced by experiments on the surface of the Earth around 1971--3, in the period when the transition from classical to quantum electrodynamics as the dominant theory was completed.
Using this standard background allows the Standard Model to define constant masses, but prevents unification
with the dynamic background required in general relativity.

In the $SO(7,3)$ model, the dynamic background at a point is quantised in terms of the photon, eight gluons and six neutrinos/antineutrinos.  Restricting to $SO(3,1)$
splits the neutrinos/antineutrinos into an electric field and a gravitational field, and splits the gluons into three antisymmetric or magnetic gluons,
and five symmetric or tidal gluons. The anti-symmetric gluons behave like tiny massless magnets, with North and South poles defining the
anti-symmetry. They therefore stick together like magnets, and form magnetic field lines, which in this model become physical objects,
though remaining massless. This may explain why magnetic field lines often behave as though they are physical objects.
The symmetric gluons behave more like a massless fluid, that flows with the tides. Again, these massless tides behave as though they are
physical objects, and have effects that may be similar to the effects of `dark matter'. That is, they gravitate, but without having (Dirac) mass.

In principle, the gluons can exist even in the absence of matter, creating large-scale invisible
structures of magnetic and tidal fields. These large-scale massless
structures then attract matter to form similar structures, that then become visible by virtue of the interactions between matter particles.

As an example of how the static and dynamic backgrounds produce different results, consider again
the mass ratios of the electron, proton and neutron. First we quantise the proton and neutron with quantum numbers $(1,1,1,1)$ and $(0,1,1,1)$
respectively, as above. Then we interpret the last three coordinates as a dependence on a direction in space, and the first coordinate 
as a dependence on time, The mass ratio of proton to neutron therefore relates to a 
tidal difference in the average vacuum state on the surface of the Earth. There are approximately 730 tides in a full rotation (i.e. a year) on Earth, which
equates to a ratio of approximately
\begin{align}
729/730 \approx .998630
\end{align}
while the experimental value of the proton/neutron mass ratio is around
\begin{align}
.998624.
\end{align}

In other words, the mass ratio obtained from experiment, using the Standard Model to define inertial mass in a constant background, 
agrees to an accuracy of 
\begin{align}
.998630/.998624 \approx 1.000006
\end{align}
that is 6ppm, with the value obtained by calculation, using the proposed new model to define gravitational mass in
terms of the tidal background. This is well within the uncertainty in experimental measurements of $G$, which is our
best experimental check on the ratio of inertial to active gravitational mass.

In the other case, we compare the electron in $(-1,1,1,0)$ with the proton in $(1,1,1,1)$, and notice that there are now two directions in space,
with an angle between them again determined by the tidal field. This angle can only be the angle between the two rotations
already considered, that is the angle between the equator and the ecliptic, also known as the tilt of the Earth's axis. If we use the sine of the average angle
of tilt in around 1971--3, that is about $23.4444^\circ$, to estimate the ratio of the electron mass to the difference between neutron and proton masses,
then we obtain a proton/electron mass ratio of about $1836.11$, which is close to the constant value $1836.15$
that the Standard Model derives from the assumption of a constant background \cite{1973}. 
However, it is evident \cite{1969} that it was not possible to reconcile all the different measurements made over the years 1949--72, as indeed
the new model predicts.
 If we use the angle of tilt $23.4368^\circ$ fifty years later, the ratio works out at
around $1836.67$. 

This means that the inertial mass defined by the Standard Model has remained fixed, while the gravitational mass, defined by the new model,
has varied with the variation of the background \cite{universal,gravimass}. 
Of course, we cannot test this prediction directly, as we cannot measure the gravitational mass of an electron.
The best we can hope for is to measure the gravitational mass of a hydrogen atom. If we assume, as is reasonable given the essentially unchanging lengths of the day and the year, that the proton and neutron 
gravitational masses
have remained constant over the period, then the gravitational mass of a hydrogen atom has changed by about 170ppb, several orders of magnitude
below what could be detected by direct experiments.

On the other hand, the calculations here only refer to an average background over a period of years. On shorter timescales it should be possible to 
detect much larger deviations of gravitational mass from the long-term average, as the background varies locally in a largely unpredictable way.
For example, direct measurements of gravitational mass of copies of the International Prototype Kilogram (IPK) will vary over time, and over space,
largely unpredictably. Experiment reveals that there are indeed variations of this type, that are large enough to be measured, and
which cannot be explained by standard physics. They are explained by the SO(7,3) model, at least qualitatively, by random unpredictable
variations in the quantised vacuum. A similar effect is visible in experiments to measure the Newtonian gravitational constant
$G$, which continue to come up with inconsistent results
\cite{Gillies,newG}.
What all these experiments reveal, therefore, is an effect of quantum gravity, that subtly alters the relationship
between inertial and gravitational mass over a sufficiently large distance in spacetime.

Indeed, any experiment that measures mass of short-lived particles using detailed information about the momentum of decay products,
including neutrinos, is potentially liable to contamination from a variable equivalence between inertial and gravitational mass, and may therefore
report different results at different times, or in different places.
The exact form of the variation is difficult to predict, and may contain a significant random component, but one should be alert
for potential correlations with the time of day, the time of year, the phase of the moon, the phase of the tides, the phase of the sunspot cycle,
the tilt of the Earth's axis, the inclination of the Moon's orbit to the ecliptic, 
the latitude of the experiment, the weather, the train timetable, the holiday season, and anything else that has an observable physical effect.

For example, the reported $W/Z$ mass anomaly \cite{WZ}
could well be attributable to such a cause, which might be the one suggested in \cite{WZ2},
that implies a dependence on (among other things) the latitude of the experiment, or it might be something else entirely.
In any case, if the dependence of the measurements on properties of neutrinos means that the result is effectively a mixture of inertial and
gravitational masses, it would be difficult to obtain a value that is constant over a period of decades, to any greater accuracy than about $10^{-4}$.

More generally, not only decay times but also decay modes may be influenced by the details of the background quantum vacuum.
Any experiment that exhibits CP-violation is likely to be susceptible to such effects, since CP-violation involves a discrepancy between
complex conjugation of charge $U(1)$ and space reversal of rotation $SO(2)$. Or, to be more precise, a discrepancy between
complex conjugation of $SU(2)$ and parity-reversal of $SO(3)$. 

The experimental fact of CP-violation is itself an argument that spin $SU(2)$
is not the same thing as the double cover of rotation $SO(3)$. 
The magnitude of the effect can be estimated from the difference between a flat complex $2$-space and a curved  real $3$-space.
The Standard Model works with a flat $SU(2)$, while gravitational effects arise from a curved $SO(3)$. The curvature of the classic experiment
\cite{CPexp} over a distance of $57$ feet amounts to about $.000002$ radians, which is just the angle required to explain the
experimental results.

What this means is that the neutral kaon eigenstates are described in this model not by $SU(2)$, as the Standard Model maintains, but by $SO(3)$.
In particular, the `quantum superposition' that is required to explain the eigenstates using the standard $SU(2)$ theory is not required
when we use $SO(3)$ instead. The kaons quickly decay, leaving only the most stable of the three (gravitational) eigenstates.
During the experiment, the kaons rotate off the eigenstate due to the curvature of the gravitational field. Hence the anomalous kaons quickly decay and restore the remaining kaons to the most stable eigenstate relative to the new background. 

Much the same thing probably happens to neutrinos. There are again three gravitational/weak eigenstates, but they do not decay, because there is nothing
lighter for them to decay into. But they still oscillate \cite{oscillation,neutrinos, SNO}
between the different eigenstates as the gravitational field curves underneath them.
In terms of experimentally measurable effects, there are two separate effects that can potentially be detected: firstly, an average effect that depends
only on the average described by the classical gravitational field, and secondly, more local and possibly random effects due to
fluctuations in, or local concentrations of, neutrino and/or gluon densities.  
In terms of theory, neutrinos come in strong eigenstates labelled by $i,j,k$, and weak eigenstates labelled by $IL,JL,KL$, in such a way that the
relationship between the two is background-dependent. The weak eigenstates are defined by the electron, muon and tau particle,
while the strong eigenstates are defined by the three directions in space.

\section{Discrete structure}
Underlying the continuous $SO(7,3)$ model there is a discrete structure of different types of elementary particles,
defined by appropriate sets of quantum numbers. There must therefore be a finite symmetry group that fits into the
structure of the $10$-dimensional representation and respects, or even explains, the symmetry-breaking
\begin {align}
10 \rightarrow 7 + 3 \rightarrow (1+2\times 3) + (2+1)
\end{align}
Since finite groups are compact, this group must fit into the $24$-dimensional compact subgroup $SO(7) \times SO(3)$.

The obvious group that fits this structure is the group $S_4$ of order $24$, that is the symmetric group of all permutations on four letters,
which is isomorphic to the rotation symmetry group of the cube..
It has five irreducible representations $\rep1^+$, $\rep1^-$, $\rep2$, $\rep3^+$ and $\rep3^-$ such that 
$\rep3^-$ describes the cube, $\rep1^++\rep3^+$ the permutation
representation on the four letters (four body diagonals of the cube), $\rep1^-$ is the sign representation that distinguishes odd from even permutations,
and $\rep1^++\rep2$ is the permutation representation on $3$ letters (three pairs of opposite faces of the cube). 
There are also monomial representations (here, these are just permutation representations with signs attached to the matrix entries) 
$\rep1^-+\rep2$ and $\rep1^-+\rep3^-$.

Then we can write the
permutation representation on the three generations as
\begin{align}
3 \rightarrow \rep2+\rep1^+
\end{align}
leaving the other representations to split
\begin{align}
7 \rightarrow \rep1^- + \rep3^++\rep3^-.
\end{align}
Alternatively, we can swap $\rep1^+$ with $\rep1^-$ and work with a more subtle monomial 
generation structure that incorporates (weak) doublets under
$\rep1^-$:
\begin{align}
3 & \rightarrow \rep2 + \rep1^-\cr
7 & \rightarrow \rep1^++\rep3^++\rep3^-
\end{align}
In either case, the splitting of the ten coordinates into four weak and six strong in the $SO(10)$ GUT implies that
the weak force is represented in $\rep1^++\rep1^-+\rep2$, and the strong force in $\rep3^++\rep3^-$,
which exhibits a `broken' weak symmetry and an unbroken strong symmetry.
The $7$ coordinates of $\rep1^\pm+\rep3^++\rep3^-$
combine part of the weak force with the strong force and can 
potentially be interpreted, not as colours, which we have implemented as continuous variables, but as $7$ different charges, for
$7$ different types of particles and anti-particles:
\begin{align}
0, \pm1/3,\pm2/3,\pm1.
\end{align}
Hence the $21$ fermionic degrees of freedom split into three generations of seven charges in this way. 

The structure of these $21$ degrees of
freedom as a representation of $S_4$ is
\begin{align}
(\rep1^\pm+\rep3^++\rep3^-) \otimes (\rep2+\rep1^\mp) = \rep1^-+\rep2+3\times\rep3^++3\times\rep3^-
\end{align}
It is unclear exactly how to interpret this decomposition, but there is some mixing between $\rep3^+$ and $\rep3^-$ arising from the
decompositions
\begin{align}
\rep2\otimes \rep3^+ & = \rep3^++\rep3^-\cr
 & = \rep2\otimes \rep3^-.
\end{align}
It is possible, therefore, that the charges emerge on the right-hand side of the equation out of the colours defined on the left-hand side. 
Or vice versa.

If we look in the same way at the $24$ bosonic degrees of freedom, we have
\begin{align}
\Lambda^2(\rep2+\rep1^\pm) & = \rep1^-+\rep2\cr
\Lambda^2(\rep1^\pm+\rep3^++\rep3^-) & = \rep1^-+\rep2+3\times\rep3^++3\times\rep3^-
\end{align}
where in the second line we see the same representation as in the fermionic case.
There therefore appears to be some kind of deep `supersymmetry' between bosons and fermions, described by the
isomorphism between these two representations. Again, the appropriate interpretation is far from clear, but $\rep1^-+\rep2$
appears in the fermionic case as three generations of neutrinos, with antineutrinos obtained by a change of sign, and in the first of the
bosonic cases it looks like a $3$-dimensional Higgs sector, obtained as the antisymmetric square of the three-generation representation.
The Higgs boson itself presumably comes from the irreducible part of this, as
\begin{align}
\Lambda^2(\rep2) = \rep1^-
\end{align}
and the other bosonic copy looks like the weak interaction, with the $Z$ boson in $\rep1^-$ and the $W^\pm$ in $\rep2$.
However, there may be a better interpretation than this.

Indeed, it may be better to abstract even further, from the representations to the group itself, where the chain of normal subgroups
\begin{align}
1 \subset V_4 \subset A_4 \subset S_4
\end{align}
of orders $1$, $4$, $12$ and $24$ respectively splits the even permutations into one identity element, $3$ elements of order $2$ and $8$
elements of order $3$, reminding us of the splitting of the bosons as $1+3+8$ in the Standard Model.
This group structure has been suggested by Jansson \cite{Jansson} as a way to describe the action of the weak force
in generation-mixing interactions, as well as allocating an efficient system of quantum numbers to the
elementary fermions. Similarly, the odd permutations
split into $6$ elements of order $2$, and $6$ of order $4$, which might describe the splitting of the $12$ fermions into $6$ leptons
and $6$ quarks. 

\section{Conclusion}
In this paper I have demonstrated that there is essentially a unique embedding of the Standard Model in $E_{8(-24)}$ and that it entails
a version of the old $SO(10)$ GUT using the real form $Spin(7,3)$. The latter can be broken down into a `gauge group'
$SU(3,1)$ plus a complex `Dirac algebra' without scalars. The complex pseudoscalars in the `Dirac algebra' 
can be interpreted as the Higgs field.
Although I have not written down a precise replacement for the Dirac equation, it is nevertheless clear that
the gauge group sits in the kernel of any putative Dirac equation of the proposed type,
so consists of particles without Dirac mass. The signature $(6,9)$ implies that there are $9$ gauge bosons ($1$ photon and $8$ gluons)
and $6$ gauge fermions ($3$ neutrinos and $3$ antineutrinos). These particles form the quantum vacuum on top of which matter sits.

I have shown how the extension from $SU(3)$ to $Spin(7)$ extends the three colours and three anticolours of the SM to a total of
$7$ `colours', or more precisely, charges. Similarly, the extension from $U(1)$ to $Spin(3)$ extends from a one-generation model
to a three-generation model, and permits some calculations of mixing angles from masses (or vice versa). These calculations are possible because
the masses defined by the Dirac equation on the quotient space $SO(7,3)/SU(3,1)$ cannot be separated from the mixing angles inside the
gauge group---all $45$ degrees of freedom are connected to each other in the simple group $Spin(7,3)$.
As a consequence, we find that the rest of $E_{8(-24)}$ is not required in the model, as $\so(7,3)$ already contains everything that exists
in the SM, except the spinors, which we no longer need, as they have been replaced by pairs of complex $4$-vectors in the Dirac equation.

Moreover, the splitting of
the Lie algebra $\su(3,1)$ into $\so(3,1)$ plus a tensor that transforms like the Einstein tensor permits a description of the vacuum that
is equivalent, to first order, to that provided by GR. However, second-order effects due to the self-interaction of the gravitational field have the
opposite sign to GR, so that it should eventually be possible to distinguish the two models experimentally.
At this point, we can look at particle physics in a dynamic rather than static background, and produce evidence that the
Standard Model is dependent on the choice of a `standard' background. 
Interactions between elementary particles and the background appear to be
responsible for many of the unexplained parameters of the Standard Model.

Of course, this model is not a `Theory of Everything', it is just a mathematical framework within which it
might be possible to develop such a theory. Significant further work will be required in order to understand the
massive particles, and how they are represented by $15$ complex dimensions of the algebra.
Complexification here first uses $U(1)$ to reverse the signature $(6,9)$ to $(9,6)$, so that there are $9$ fundamental
massive fermions and $6$ fundamental massive bosons. The masses themselves are added in using the real scalar $SO(1,1)$.
But to make more progress here
we need an explicit replacement for the Dirac equation, that allows us to calculate the masses of elementary particles from a combination of the
particle quantum numbers and the properties of the dynamic quantum background.

Many other mathematical frameworks for unification have been proposed, but what our analysis shows is that it is very
important to keep tight control on the correct real and complex forms of all the groups and algebras that are involved. 
In particular, a framework based on $SO(10)$ can only work in the signature $SO(7,3)$, in order to get rid of the unwanted gauge bosons that mediate non-existent forces, and replace them with gauge fermions (neutrinos and antineutrinos), that are responsible for infinitesimal changes to the relationship between
space and time, momentum and energy, and mass.
It follows that a fundamental requirement for a unified theory of gravity and particle physics is to generalise the Yang--Mills paradigm to
noncompact groups, and hence allow gauge fermions into the theory. 

Frameworks based on other subgroups of $SO(7,3)$ may also be viable, provided they are interpreted correctly. For example,
a subgroup $SO(7) \times SO(3)$ might be implemented in terms of imaginary octonions together with imaginary quaternions, leading to a compact
gauge group of dimension $24$ and $21$ non-compact dimensions in the tensor product of imaginary octonions with imaginary quaternions.
This idea is the basis for models developed by Dixon \cite{Dixon1,Dixon2}, 
and more recently by Furey \cite{Furey1,Furey2}. A subgroup $SO(3,1)\times SO(4,2)$ supports a `twistor' model 
which combines a `left-handed' Weyl spinor and a `right-handed' Penrose twistor, as proposed by Woit \cite{WoitSO24,WoitRH}.

\end{document}